\documentclass[manuscript]{aastex}
\slugcomment{}
\shorttitle{Mrk~231}
\shortauthors{Low et al.}
\slugcomment{Draft version:(\today)}

\begin{document}

\title{High Spatial Resolution HST/NICMOS Observations of Markarian~231}
\author{F.~J.~Low,~G.~Schneider,~G.~Neugebauer}
\affil{Steward Observatory, University of Arizona}
\affil{Tucson, AZ 85721}
\email{flow@as.arizona.edu,gschneider@as.arizona.edu,gxn@as.arizona.edu}

\begin{abstract}{
Observations of Markarian~231 at 1.1~$\mu$m taken with NICMOS 
on the Hubble Space Telescope are described.\ The brightness of
the object in the near infrared and the inherent short-term
stability of the NICMOS optical and instrumental system enables
application of special observational and analysis techniques
that effectively increase high spatial resolution.\  By these
means, we set an upper limit on the size of the core of the AGN
nucleus  at 8~mas corresponding to a radial projected distance
from the center of Markarian~231 of $\sim$~3~pc.  }

\end{abstract}

\newpage
\section{Introduction}
The ultra-luminous infrared galaxy (ULIRG;
luminosity~$>$~10$^{12}~$L$_\sun$) Markarian~231 (Mrk~231), with
a bolometric luminosity of 3.3~$\times$~10$^{12}~$L$_\sun$, was
first identified as a ULIRG by Rieke~\&~Low~(1972) and was found
by the IRAS survey to be the most luminous object within 300~Mpc
(Soifer et al.\ 1986).\ The ratio of its flux density at
25~$\mu$m~(f$_{\nu}$(25)) to its flux density at
60~$\mu$m~(f$_{\nu}$(60))
---~f$_{\nu}$(25)/f$_{\nu}$(60)~$=$~0.3~--- leads to its
inclusion as a  $``$warm$"$ IRAS galaxy (Low et al.\ 1988). It
has the visual spectrum of a Seyfert~1 galaxy and a redshift
z~$=$~0.042 implying a projected distance scale of 800~pc~asec$^{-1}$.
From the 2MASS survey, the near  infrared magnitudes within a
7$''$ diameter beam of Mrk~231 are J(1.25~$\mu$m)~$=$~11.02~mag,
H(1.65~$\mu$m)~$=$~10.04~mag, and
K$_S$(2.15~$\mu$m)~$=$~8.94~mag (Skrutskie et al., 1997). We
take the Hubble constant to be
H$_0$~$=$~75~km~s$^{-1}$~Mpc$^{-1}$.

Because Mrk~231 has been  considered to be a nearby  infrared
quasar, several observations over a range of wavelengths have
been undertaken to resolve its core.  Lai et al.\ (1998), using
adaptive optics with the 3.6-m diameter CFHT Telescope, set a
limit on the size of the full width at half maximum (FWHM) of
the core of the  active galactic nucleus (AGN) of 0$\farcs$11
(90~pc) in the near-infrared. \  Quillen et al.\ (2001) included
Mrk~231 in a survey   of unresolved continuum sources at
1.6~$\mu$m with the Hubble Space Telescope (HST); they set a
limit on the FWHM of Mrk~231 of $\leq$~0$\farcs$13 (100~pc).\ 
Soifer et al.\ (2000), using the 10-m diameter Keck Telescope,
set a limit on the size of the AGN core  of 0$\farcs$13 (100~pc)
at 12.5~$\mu$m. Kl\"{o}ckner, Baan \& Garrett (2003) report that
the hydroxyl (mega)-maser emission shows the characteristics of
a rotating, dusty, molecular torus (or thick disk) located
between 30 and 100~pc from the central engine of Mrk~231.
Lonsdale et al.\ (2003) showed, from VLBI continuum imaging
observations, that at 18~cm wavelength Mrk~231 consists of a
single core of FWHM $<$ 0$\farcs$005 (4~pc) plus a 0$\farcs$03 
extension to the south. Further discussion of the galaxy  and
references emphasizing other important aspects of the
observations of  Mrk~231 are given, e.g., in Lonsdale et al. and
Smith et al.\ (1995).

The capability of using the HST at near-infrared wavelengths,
its short-term (intra-orbit) optical stability (Schneider et
al.\ 2001), and the extreme brightness of Mrk~231  offers a
singular opportunity to set physically interesting limits on the
size of the core in the near-infrared. In this paper  we present
observations of Mrk~231 made with  NICMOS on the HST that were
especially designed  and optimized to probe the point-like
source in Mrk~231.  

On the basis of its far infrared color, Hines et al.\ (2004)
include Mrk~231 as one of nine hyperluminous and $``$warm$"$
ultraluminous infrared galaxies observed with NICMOS. Although
the observations of Mrk~231 in Hines et al.\ are the same as
shown here, the comparison star used by Hines et al. to define
the point spread function (PSF) differs from the one used here
and their reduction of the image is a preliminary one. The
papers by Hines et al. and by Quillen et al.\ (2001) both
include Mrk~231 as one of a larger sample and hence do not
attach special attention to the reduction of its data. In this
paper we have concentrated on the reduction of this one object
and have attempted to achieve the highest spatial resolution
possible with these data. 

\section{Observations}
Observations of Mrk~231 were made in September, 1998 using the
NIC1 camera and the F110M filter which has a central wavelength
of 1.1~$\mu$m and filter  full width at half maximum 
of 0.2~$\mu$m as described by Thompson et al.\ (1998)  and in the
NICMOS Instrument Handbook (Roye et al., 2003).\  The camera
has  0$\farcs043$~$\times$~0$\farcs043$ pixels corresponding to
the diffraction limit of the HST at 1.0~$\mu$m, and the  array
has 256~$\times$~256 pixels so the field of view is
$\sim$~11$''$~$\times$~11$''$. Observations at 1.6 and
2.1~$\mu$m using the NIC2 camera accompanied the 1.1~$\mu$m
observations; these and related observations will be described
in Schneider \& Low\ (2005).
 
Four sequential, but separate, images of Mrk~231 were obtained.\
The observations were made using STEP1/NSAMP=24 multiaccum
sampling (Roye et al., 2003), so the total integration time at
the end of the 24-read exposure ramp for each of the individual
exposures in the four point pattern was 22 seconds.\ After each
exposure, the telescope was offset by a nominal 50.5 pixels so
the active nucleus was centered on different relative positions
of a pixel as well as sampling different portions of the array.
In fact the  observed offsets differed slightly from half
integer pixels since the pixels projected onto the sky are
slightly rectangular and the average offset was $\sim$~50.6
pixels.  

Four separate images of the  Hubble guide star GSC~0384500748 
at right ascension 12$^{h}$~56$^{m}$~15.8$^{s}$, declination
$+$56$\degr$~48$^{'}$~17$\farcs$7~(J2000) ---~a K star~--- were
similarly obtained in the orbit immediately after the Mrk~231 
observations to determine the PSF using the same offset pattern
as had been used with Mrk~231.\  There is no field dependency in
the structure of the NIC1 camera PSF star image.\ The PSF star 
was selected primarily due to its close proximity
---~$\sim$~4$'$~--- to Mrk~231 in the sky. The 2MASS survey
magnitudes of the PSF star are J~$=$~11.13~mag, H~$=$~10.62~mag
and K$_S=$~10.54~mag (Skrutskie et al., 1997) so it is of 
comparable brightness to Mrk~231 in the J band although fainter
at 2.2~$\mu$m and considerably less red in its H-K color.\ Both
sets of observations were carried out at the same spacecraft
roll  angle, i.e., the same field orientation, to minimize
changes in differential heating to the optical telescope
assembly, and better stabilize the PSF.  The time and roll
constraints, and the very close proximity of the PSF star in the
sky, ensured the PSF was stable between the Mrk~231 and
reference observations.

\section{General Results and Common Reduction} 
The images of Mrk~231 and of the PSF star were  first combined
to make separate images each containing a single image of either
Mrk~231 or the PSF star following the precepts outlined in
Schneider \& Stobie (2002).\ Image centroids were determined by
Gaussian profile fitting after re-sampling the images onto 32
times  finer grids via bi-cubic interpolation
apodized by a sync function.\ Once the centroids were
determined all images were co-aligned by shifting three of the
images to the location of the fourth to the nearest integral
pixel and then by interpolative sub-pixel re-sampling also by
bi-cubic sinc-apodized interpolation. The four images of the PSF
star, which were  originally sampled according to the Nyquist
criterion with $\sim$ half pixel offsets,  were then median
combined after registration providing super-critical sampling of
the PSF star. The final image was re-sampled onto a grid of 
$0\farcs$0054  pixels.

The resultant image of Mrk~231 is shown as the left panel of
Figure~1. The  image of the PSF star  is included as the right
panel of Figure~1 scaled in intensity to ``match" the nuclear
region of Mrk~231.\ In the region with radius~$\leq$~0$\farcs$2,
the sum of the squares of the subtraction residuals in the
difference image was minimized along with the the total energy
in the difference image.\ The imperfectly pupil-plane masked
NICMOS+HST ``diffraction spikes" in the region
radius~$>$~0$\farcs$2 were simultaneously minimized.\ See
Schneider et al.\ (2001) and Krist et al.\ (1998) for the
details of HST/NICMOS target:reference PSF scaling and the
effects of mis-matched PSF image structures arising from 
thermal de-spacing of the HST secondary mirror.\ The effects of
temporal variability in pupil mask alignments are also
discussed.\ Figure~2 shows  median combined azimuthal radial profiles
of Mrk~231 and of the PSF star and their integrated curves of
growth.\ It is clear the nuclear region of the 1.1~$\mu$m
Mrk~231 image is essentially that of an unresolved point source;
the FWHM of the raw image of Mrk~231 and of the PSF star is
$\sim$~0$\farcs$10.\ In both Figure~1 and Figure~2,  Mrk~231
shows the presence of a host galaxy seen at low surface
brightness at radius~$>$~0$\farcs$2, but this is better examined
by much deeper corongraphic imagery  and will be discussed in
depth by Schneider \& Low (2005). 

\subsection {Intrinsic Core Size Limit}

To recapitulate, the raw 0$\farcs$043~$\times$~0$\farcs$043
pixels of NIC1 were over-sampled at 1.1~$\mu$m  through the half
integer pixel offset and the high signal to noise ratio
available on Mrk~231 to effectively act as
0$\farcs$022~$\times$~0$\farcs$022 pixels.\ With such well 
sampled images it was then possible to set a limit on the FWHM
of a putative extension present in the nucleus of Mrk~231 well
below the classical diffraction limit of the telescope by
comparing the image structure of the Mrk~231 nucleus to that of
a model made up of a core of negligible angular extent plus an
assumed extension. We found that radial profiles of PSF
subtracted images of various comparisons were superior in
delineating the size limit of the core to direct techniques,
e.g., a Richardson-Lucy deconvolution.

The left image of Figure~3 shows the result of subtracting the
observed image of the PSF star, after registration and
flux-scaling, from the Mrk~231 image.\ Images constructed  by
convolving the observed image of the PSF star with a model of
Mrk~231 in which Mrk~231 is assumed to follow a Gaussian profile
of 5.4 and 10.8~mas FWHM (middle and right) are included in
Figure~3.\ We take the similar features seen in the middle and
right hand images of Figure~3 to be a signature of the putative 
broadening in the model of Mrk~231. 

Figure~4 shows median combined azimuthal radial profiles of the PSF
subtracted images of Figure~3. \ The three solid curves in the
figure refer to  putative extensions for the core of Mrk~231,
again after subtracting the observed PSF star.\ The FWHM
parameter of the Gaussian profile assumed for Mrk~231 is given
in the figure.
 
In order to test the influence of noise in the image of the PSF
star,  the  image of the observed PSF star was subtracted from a
noiseless image of a model NICMOS PSF
star$\footnote[1]{http://www.stsci.edu/software/tinytim/tinytim.html}$,
as shown by the dotted line in Figure~4.\ The curves for finite
Gaussian extensions  are virtually the same whether the image of
the noiseless model PSF or observed PSF stars are subtracted.\
Additionally, the image of Mrk~231 itself was used in the
subtraction; no significant difference in the final difference
image was seen.

The negative residual which arises when normalizing the raw data
to the central pixel is not surprising.\  This
zonal under-subtraction is, in fact, replicated using either the
observed PSF star (dark solid line in Figure~4) or a noiseless
model PSF (dotted line in Figure~4), which suggests this residual is real
and not an artifact of a mis-matched PSF.\ It is consistent with
the amount of differential de-spacing of the HST secondary mirror
expected between the Mrk~231 and PSF star observations.\ This residual cannot
be explained by extended flux with a Gaussian radial brightness
profile as it would non-physically imply that the point-source
reference PSF star is wider than the core of Mrk~231.  

The signature emission seen in Figure~3 is very clearly evident
in Figure~4 as a non-zero emission peak at a radial distance of
$\sim$~0$\farcs$075 --- a minimum in the NICMOS diffraction
pattern --- in the difference image; this  can be used to set a limit
on the putative extent of the core of Mrk~231.\ It is clear from Figure~4
that a Gaussian source with a FWHM of 10~mas   would be
unequivocally detected in the difference image, while a Gaussian
profile with a FWHM of 5~mas could be hidden in the data. We
will take a FWHM of 8~mas as the limit for the putative
extension of Mrk~231.

\section{Discussion}
\subsection {Core Size}
In order to place the size limit deduced from these observations
in context, it is useful to consider  a simple
equilibrium thermal model of dust grains surrounding, and in
thermal equilibrium with, a luminous hot central source. At
1.1~$\mu$m, the maximum blackbody emission ($\nu$f$_{\nu}$)
occurs at a temperature $\sim$3000~K. Dust grains, however,
sublimate at a temperature $\sim$2000~K so a temperature of
2000~K for the dust grains will be assumed. If the central
source has a bolometric luminosity
3.3~$\times$~10$^{12}$~L$_{\sun}$, black-body grains 0.17~pc
from the central source come to equilibrium at  this
temperature. Silicate grains about six times further out, or
1~pc from the central source, come to equilibrium.  The size
limit we have deduced from  these observations and NICMOS is a
Gaussian FWHM of 8~mas or a projected radial distance of
3~pc at 1.1~$\mu$m.\ Although this is significantly smaller 
than previously published near-infrared limits, this does not
restrict physically possible models.

\acknowledgments{}  
We thank Paul S. Smith and Dean Hines for help with the
manuscript and source selection. We also thank an anonymous
referee for valuable suggestions.\ This work was supported, in
part, by NASA grant NAG  5-3042 and 10843 to the NICMOS
Instrument Definition and Guaranteed Time Observing teams.  This
paper is based upon observations with the NASA/ESA Hubble Space
Telescope, operated by the Space Telescope Science Institute,
which is operated by the Association of Universities for
Research in Astronomy, Inc., under NASA contract NAS 5-26555.

\newpage
\figurenum{1}
\figcaption{
NICMOS Camera NIC1 F110M images are given of Mrk~231 (left) and
of the PSF reference star GSC~0384500748 (right) with the flux
density renormalized as discussed in the text.
Both images are 7$\farcs$5 ~$\times$~7$\farcs$5 and have been
constructed from four-point half-integral pixel stepped
observations, as described in the text, providing effective
spatial sampling of 0$\farcs$022 per re-sampled pixel. Images
are shown with a logarithmic stretch of [-2] to [+1] dex
ADU~sec$^{-1}$~pixel$^{-1}$. The diffuse brightening in the circumnuclear
region beyond the display-saturated first Airy ring in Mrk~231
compared to GSC~0384500748 is due to the presence of the host
galaxy and is discussed in Schneider \& Low (2005). 
}

\figurenum{2}
\figcaption{ 
The azimuthally median filtered surface brightness radial
profiles (top) and the integrated flux densities in appropriate
annuli (bottom)
of Mrk~231 and the reference PSF star are shown.\ The PSF star
has been scaled as described in the text and  in Schneider et
al.\ (2001) and Krist et al.\ (1998).\ It is seen that the
nuclear cores of the two objects are  extremely well matched at
radii r~$<$~0$\farcs$2.  At r~$>$~0$\farcs$2 the contribution to
the surface brightness due to the host galaxy is readily
apparent (also see Figure~1).
}

\figurenum{3}
\figcaption{ 
Images of Mrk~231 are shown after subtracting the flux-scaled
and concentricity registered reference PSF star image as
discussed in text --~left: Mrk~231 as observed--~middle: Mrk~231
assumed to follow a 5.4~mas FWHM Gaussian profile --~right:
Mrk~231 taken to have a 10.8~mas FWHM Gaussian profile. Images
are stretched from -2\% (black) to +7\% (white) of the
un-subtracted peak intensity of the Mrk~231 nucleus. Images are
0$\farcs$460 $\times$  0$\farcs$460.
} 
                                                                         
\figurenum{4}
\figcaption{                                                                                         
Radial brightness profiles (azimuthal median of all pixels in a
5.4~mas wide annuli) of the PSF subtracted Mrk~231 images shown
in Figure~3 are given.\ As discussed in the text, the solid
lines refer to the FWHM of Gaussian profiles assumed for the
core of Mrk~231.\ The  curve marked ``0~mas" represents the
image of the observed PSF star. The  subtraction of a noiseless
high fidelity model PSF, rather than the image of the observed
PSF star, from the observed Mrk~231 image is shown by the dotted
line. Error bars represent standard deviations (1-$\sigma$) of
all pixels in each annulus about the median value in each
annulus.        
}


\newpage 

\begin{references}{
Hines, D. C., Low, F. J., Evans, A. S., Scoville, N. Z., \&
Schneider, G. 2004, in preparation

Kl\"{o}ckner, H-R., Baan, W. A. \& Garrett, M. A. 2003, Nature,
421, 6925

Krist, John E., Golimowski, David A., Schroeder,~Daniel~J., \&
Henry,~Todd~J. 1998, PASP, 110, 1046


Lai, O., Rouan, D., Rigaut, F., Arsenault, R., \& Gendron, E.
1998, A\&A, 334, 783

Lonsdale, C. J., Lonsdale, C. J., Smith, H. E., \& Diamond, P. J.
2003, ApJ, 592, 804

Low, F. J., Cutri, R. M., Huchra, J. P., \& Kleinmann, S. G.
1988, ApJ, 327, L41

Quillen, A. C., et al. 2001, ApJ, 547, 129

Rieke, G. H., \& Low, F. J. 1972, ApJ, 176, L95

Roye, E. \& Knoll, K., et al. 2003, NICMOS Instrument  Handbook,
Vol. version 6.0 (Baltimore: STScI)

Schneider,~G., Becklin,~E.~E., Smith,~B.~A., 
Weinberger,~A.~J., Silverstone,~M., \& Hines,~D. 2001, AJ, 121, 525

Schneider, G. \& Low, F. J. (2005) in preparation

Schneider, G., \& Stobie, E. 2002. in Astronomical Data Analysis
Software and Systems XI, Pushing the Envelope: Unleashing the
Potential of High Contrast Imaging with HST, eds. D. A.
Bohlender, D. Durand, \& T. H. Handley (San Francisco:
Astronomical Society of the Pacific), 382

Skrutskie, M. F., et al. 1997. in The Impact of Large Scale
Near-IR Sky Surveys, The Two Micron All Sky Survey (2MASS):
Overview and Status., eds. F. Garzon, \& e. al. (Dordrecht:
Kluwer Academic Publishing Company), 25

Smith, Paul S., Schmidt, Gary D., Allen, Richard G., \& Angel, J.
R. P. 1995, ApJ, 444, 146


Soifer, B. T., et al. 2000, AJ, 119, 509

---. 1986, ApJ, 303, L41

Thompson, R., Rieke, M., Schneider, G., Hines, D., \& Corbin, M.
1998, ApJ, 492, L95

}
\end{references}
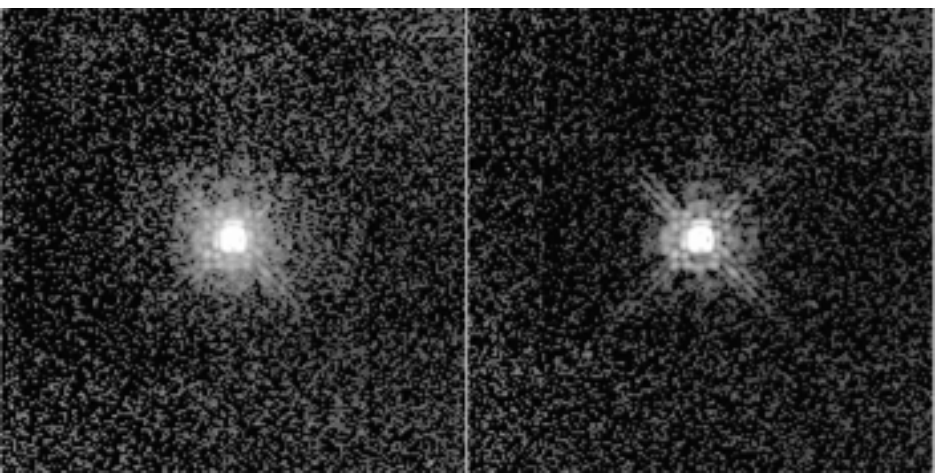
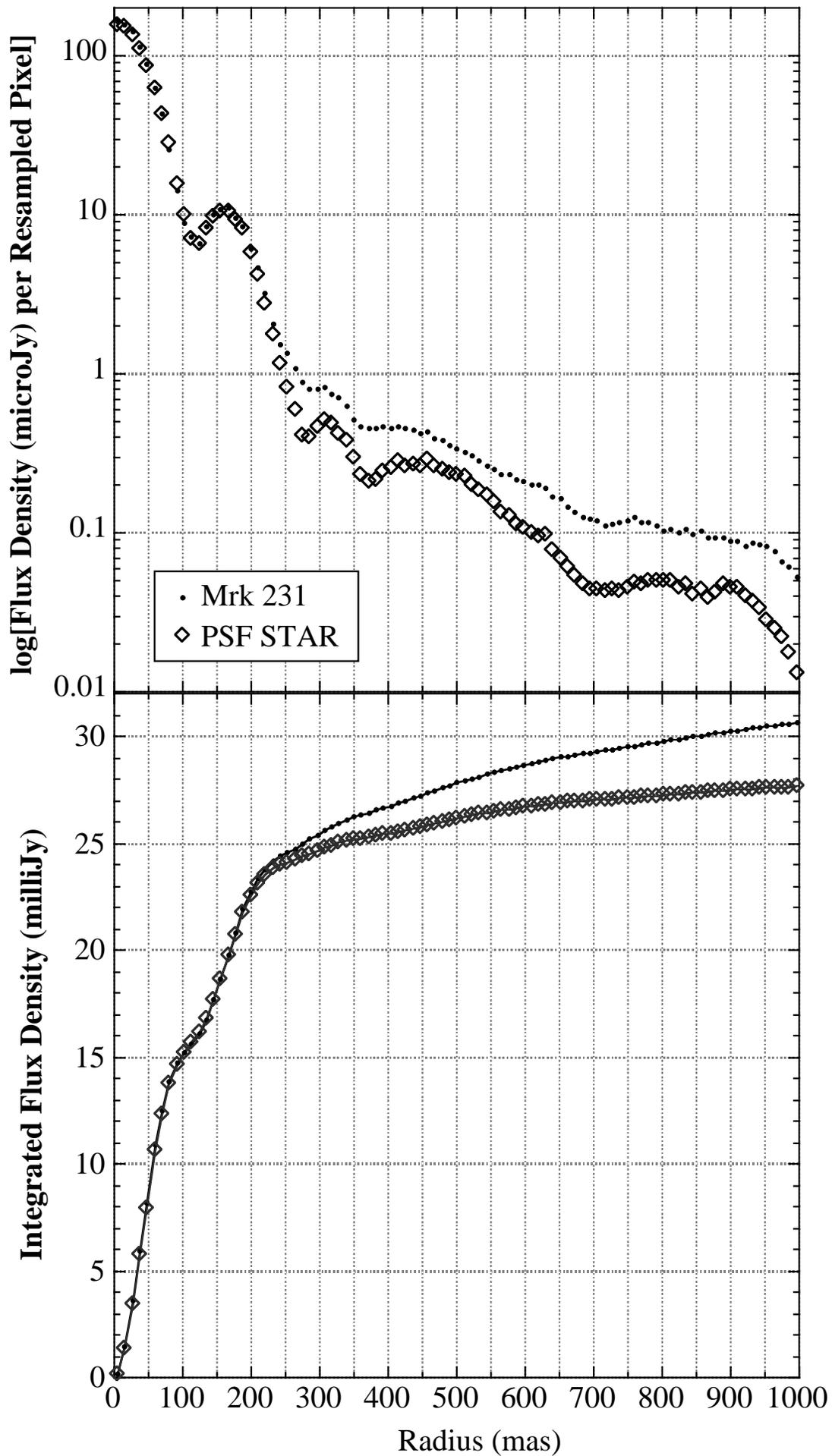
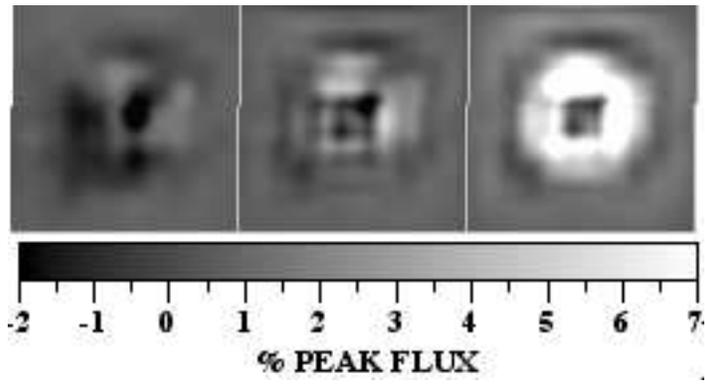
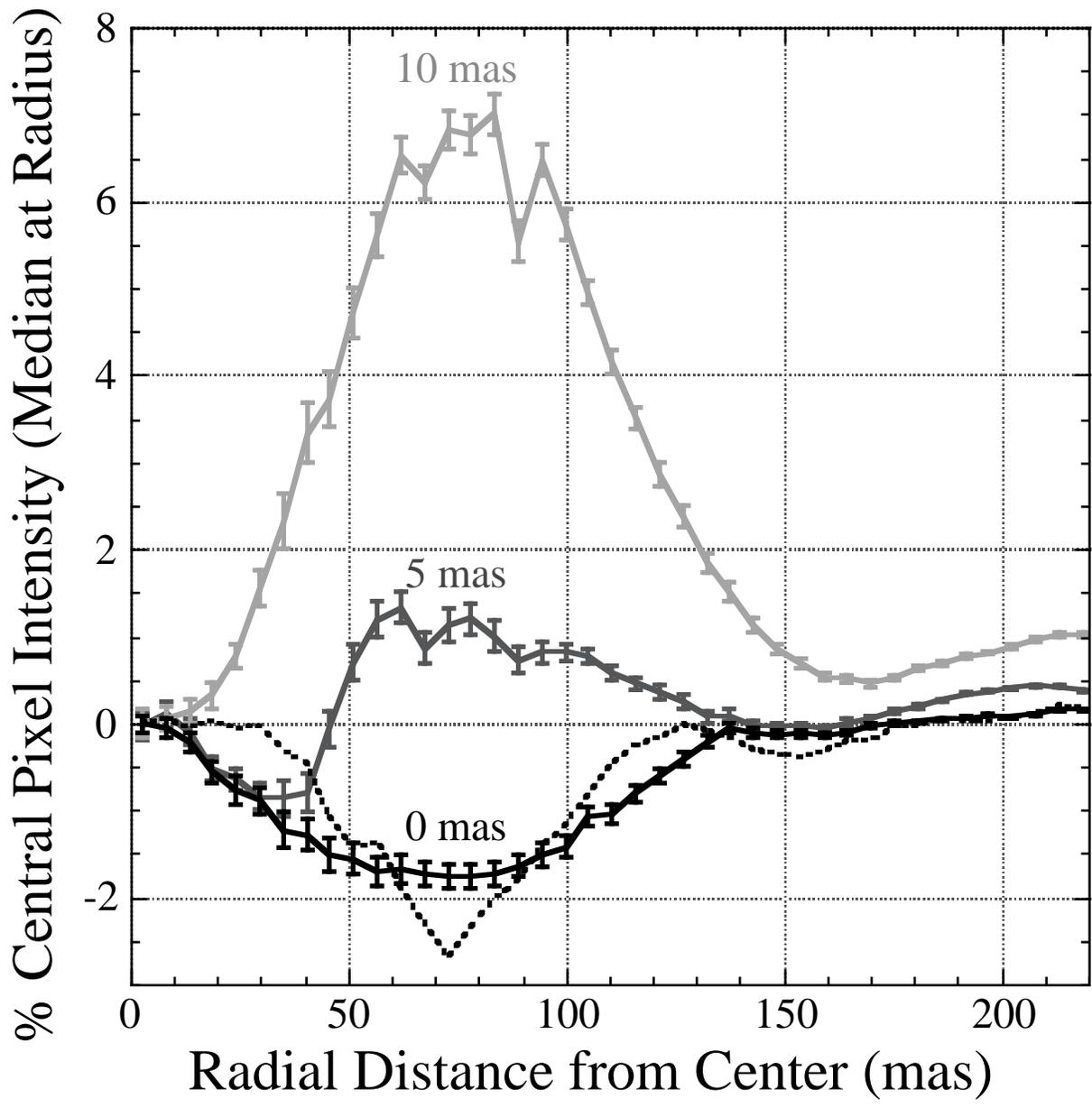
\end{document}